\documentclass[twocolumn,showpacs,preprintnumbers,amsmath,amssymb,pra]{revtex4}
\usepackage{graphicx}

\begin{document}
\title{Realization of a semiconductor-based cavity soliton laser}

\author{Y. Tanguy}
\author{T. Ackemann}
\email[]{thorsten.ackemann@strath.ac.uk}\author{W. J. Firth}
 \affiliation{SUPA,
Department of Physics, University of Strathclyde, 107 Rottenrow,
Glasgow G4 ONG, Scotland, UK}
\author{R. J{\"a}ger}
\affiliation{ULM Photonics GmbH, Lise-Meitner-Str.~13, 89081 Ulm,
Germany}

\begin{abstract}
The realization of a cavity soliton laser using a vertical-cavity
surface-emitting semiconductor gain structure coupled to an external
cavity with a frequency-selective element is reported. All-optical
control of bistable solitonic emission states representing small
microlasers is demonstrated by injection of an external beam. The
control scheme is phase-insensitive and hence expected to be robust
for all-optical processing applications. The motility
of these structures is also demonstrated.
\end{abstract}


\maketitle

Self-localized structures in driven non-equilibrium systems, loosely
termed dissipative solitons, have attracted great interest because
of their importance in a wide variety of fields (see e.g.
\cite{akhmediev05} for a recent review). They are particularly
interesting in optics -- where they are usually referred to as
cavity solitons (CS) -- because of potential applications to the
all-optical control of light,   a major thrust of modern photonics
(e.g.\ \cite{cotter99}). While some {\em laser} schemes have been
proposed and/or demonstrated to sustain CS
\cite{bazhenov92,saffman94a,taranenko97,rosanov02,bache05}, no
viable {\em cavity soliton laser (CSL)} has been developed for a
major laser technology. This useful device would convert broad-area
excitation into a narrow, coherent, power beam of high quality, or
into a controllable number of such beams.  Here we demonstrate
optically-controlled excitation and erasure of tiny self-localized
`lasers' within a semiconductor vertical-cavity surface-emitting
laser (VCSEL) structure. Our scheme is easily extensible to all
types of VCSELs, and  is therefore in the mainstream of laser
engineering. It could have a substantial impact, both as a
fundamentally-new type of laser and in the numerous technologies 
reliant on semiconductor lasers. Furthermore,
from a fundamental point of view, the optical phase in an
incoherently-pumped CS laser is an extra degree of freedom, which
presents new opportunities for fundamental studies, e.g.\ of
dynamics and interaction properties of the individual CS
\cite{rosanov05}. It will be interesting to compare the effects of
phase-sensitive interactions of laser cavity solitons with the
wealth of phenomena known for propagating spatial solitons
\cite{stegeman99}.

The achievement, control and understanding of CS have shown
remarkable progress in recent years, see, e.g.,  \cite{lugiato03,
barland02} and Chapters 3-6 of \cite{akhmediev05}. So far, however,
nearly all realized schemes have relied on driving by a broad-area
holding beam of high spatial and temporal coherence. This holds in
particular for semiconductor microcavities, which are ideal for
photonics applications because of compactness, speed and ease of
integration \cite{barland02} and where CS have been observed in a
variety of configurations. Below laser threshold, these include both
passive (i.e. absorbing, \cite{taranenko01}) and active (i.e.
amplifying, \cite{barland02,barbay06}) media. Above threshold, CS
have been demonstrated in lasers with injected signal
\cite{larionova05,hachair06}. In all these cases, the frequency and
phase of the CS is locked to that of the injected field, and so is
sensitive to any phase variations or fluctuations. This enables
efficient writing, erasing and manipulation of CS \cite{hachair05}.
It also requires, however,  active phase control within and between
devices, and this is a drawback for applications. Hence there are
advantages in removing the need for a holding beam and for
phase-controlled addressing by moving to a lasing configuration with
self-sustained CS. This also means that the device can draw its
energy from an inexpensive, incoherent source (dc current supply in
our case). Our realized {\em cavity soliton laser} has these
important properties.

\begin{figure}[ht!]
\centerline{
\includegraphics[scale=0.1,angle=0]{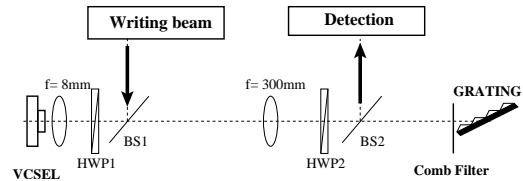}
     }
\caption{Experimental apparatus. The external cavity is $616$~mm
long, including an $8$~mm aspherical lens used as a collimator and a
$300$~mm lens: it is thus self-imaging, projecting a 37.5x image of
the VCSEL onto a Littrow-mounted holographic diffraction grating
($1800$~lines/mm). Beam sampler BS1 couples a writing beam into the
cavity, while BS2 couples out part of the beam for detection.
Half-wave plates (HWP) are used to match the principal axes of the VCSEL,
the intra-cavity beam splitters and the grating. The strong
polarization dependence of the grating's diffraction efficiency
ensures that the intracavity field has a well-defined linear
polarization. Hence the HWPs can be also used to control the
coupling efficiency of the writing beam and of the detection arm to
the external cavity. For some measurements a comb filter is added,
in the re-imaged near-field close to the grating. Its slits are
orthogonal to the grating lines.} \label{fig:setup}
\end{figure}

Cavity soliton laser schemes reported previously used either dye or
photorefractive gain media \cite{bazhenov92,saffman94a,taranenko97}
and employed intra-cavity saturable absorbers to favor bistability
-- a positive indicator for CS \cite{lugiato03,akhmediev05} --
between the lasing and non-lasing states. A semiconductor-based CSL 
would be much faster, more compact and more reliable than these systems. 
A recent theoretical paper considered a semiconductor CSL using a 
saturable absorber \cite{bache05}. We adopt a different approach, 
based on using a VCSEL in conjunction with a frequency-selective 
external cavity, which is attractive as it can  be implemented 
with essentially any VCSEL structure, using off-the-shelf optical 
components. Encouragingly, previous
experiments have demonstrated bistability in small-area VCSELs in
similar feedback configurations \cite{naumenko06,tanguy06b}. The
mechanism for bistability was shown to be due to phase-amplitude
coupling \cite{henry82}, by which the different  carrier densities
of the lasing and non-lasing states imply different refractive
index, and hence different cavity resonance frequencies. There can
thus be a stable non-lasing state, with the cavity and external
(grating-controlled) frequency out of resonance, coexistent with
stable in-resonance lasing \cite{naumenko06}. Self-localization of
the lasing state to form a CS sitting on a non-lasing background
requires a nonlinear transverse effect to sustain "gap" states below
the band of extended lasing states. Self-focusing in a broad-area
VCSEL can supply such an effect. Indeed, CS have been found in a
rather similar model system \cite{paulau07}.

The experimental set-up is shown in Figure~\ref{fig:setup}. The
VCSEL used is a broad-area bottom-emitting device, emitting at
$980$~nm and electrically pumped through a $200~\mu$m diameter
circular oxide aperture. The epitaxial structure is similar to the
one described in~\cite{grabherr99,barland02}. The self-imaging
external cavity includes two lenses and a holographic grating in
Littrow configuration. Due to the self-imaging geometry, the high
Fresnel number of the solitary VCSEL, and thus the potential for
self-localization independent of the boundaries, is preserved. Two
beam samplers are also added, one to couple out a fraction of the
beam for measurements, and the other to inject a narrow writing beam
(WB) from a tunable source.

    %The Littrow set-up was chosen for simplicity. Using the grating at
    %grazing incidence in a Littman configuration was checked to work
    %also.   However,  the resulting setup is more complex, involving two
    %2-lens telescopes in the external cavity.

The external cavity can be analyzed with the formalism of ABCDEF
matrices~\cite{martinez88}, which allows for optical elements with
angular dispersion. In our Littrow configuration, the angular
dispersion was calculated to be 0.46~mrad/GHz, while 
the angular width of the VCSEL resonance is estimated from the Bragg mirror properties to be 26~mrad.  Hence we estimate the bandwidth of
the feedback in frequency space to be about $55$~GHz.

\begin{figure}[ht!]
\centerline{
\includegraphics[scale=0.5,angle=0]{Figure2a.eps}
\includegraphics[scale=0.8,angle=0]{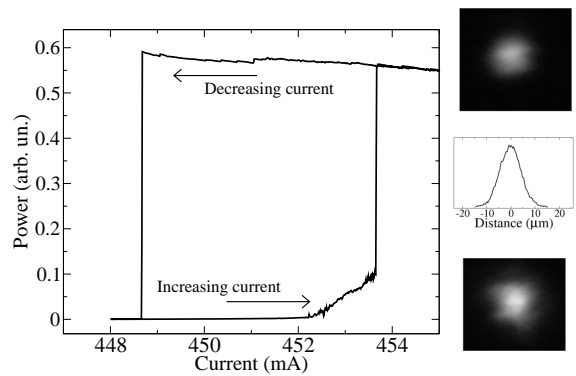}
     }
\caption{Power versus current for a single spot. In this case the
comb filter (explained below) was inserted in order to reduce the
optical background of the spot and hence obtain a cleaner
hysteresis. The panels on the right show, from top to bottom: the
near-field intensity distribution around the spot, a transverse
intensity profile through the center of the near-field distribution,
the far-field intensity distribution. } \label{fig:LI}
\end{figure}

CS are self-localized bistable structures in spatially-extended 
dissipative systems. This prescription is easy to check in
theoretical modesl, but CD identification in necessarily imperfect
experimental schemes is non-trivial. The established
procedure (e.g.\ \cite{barland02}) is to demonstrate (i)
existence at different locations in the transverse plane  of 
localized bistable structures with a well-defined shape and (ii) 
their mutual independence (e.g. by independent switch-on and switch-off).
A further indicator (iii) is then `motility', i.e. that they can 
be easily moved around by perturbations, and hence are
independent of boundary conditions and self-localized. In order to
claim a CSL, the transition (iv) to narrow-band
coherent emission obviously needs to be shown also.

We demonstrate fulfilment of criteria (i) and (iv) by
studying the behavior of the near-field under variation of the injected
current. The VCSEL is biased below the threshold of the solitary
laser, and the grating is aligned so that its frequency of maximal
feedback is red-detuned with respect to the longitudinal resonance
frequency of the VCSEL. Increasing the injection current, we observe
spontaneous formation of several localized spots, all of similar size and brightness, with a diameter of
about $10~\mu$m (FWHM; see the upper right panels in
Fig.~\ref{fig:LI}). These spots display bistability in dependence on
current, with abrupt switch on and off (see Fig.~\ref{fig:LI}, left
panel). Measurements with a scanning Fabry-Perot interferometer
indicate that a spot can, depending on parameters such as current
and temperature, operate on one or more longitudinal modes of the
external cavity (which are separated by about 240~MHz). The
single-mode emission  linewidth is $10$~MHz, i.e., each of these
spots is a tiny laser emitting coherent light.

\begin{figure}[th!]
\centerline{
   \includegraphics[scale=0.95,angle=0]{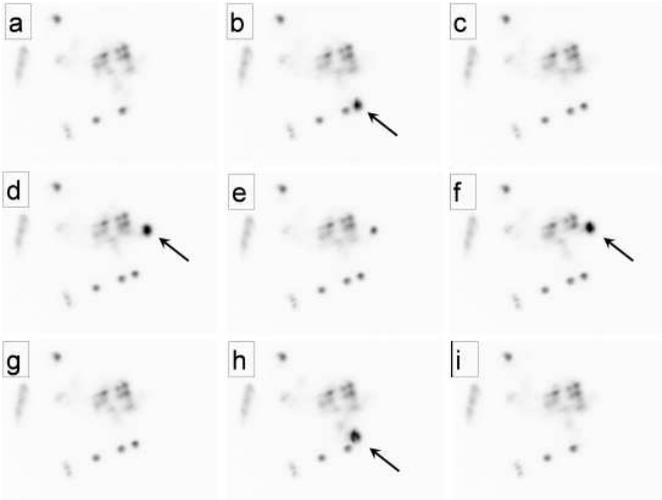}}
\caption{Near-fields showing the successive switching on and off of
two spots with an injected incoherent field (brightest spots,
indicated by arrows). Dark areas correspond to high intensities. The
writing beam (WB) is derived from a tunable laser source, with
wavelength tuned in the vicinity of the VCSEL cavity resonance. It
is focused onto the VCSEL with a $12~\mu$m spot diameter (FWHM) and
a power in the mW range. Two sites where spontaneous spots could be
observed were selected for WB injection, and the VCSEL was biased
within their bistability range. a) Both spots are off, b) injection
of WB, c) one spot is switched on and remains after the WB is
blocked, d) injection of WB at second location, e) second spot
remains on, f) WB injected beside second spot, g) second spot
switched off and does not reappear (first spot unaffected), h)
injection of writing beam to switch off first spot, i) both spots
remain off. } \label{fig:nf_switchONOFF}

\end{figure}

Regarding (ii), independence and simultaneous bistability
of two of these spot-lasers is demonstrated by
independent switch-on and switch-off by an injected field. The
sequence is shown in Fig.~\ref{fig:nf_switchONOFF}. The WB is
incoherent with the spots, so that the switching is
phase-insensitive. The mechanism is discussed below. We demonstrate
(iii), motility of these microlasers, by perturbing
them with the WB injected at some distance from the spot. They
move towards it, and can be dragged around with it for
significant distances (several beam diameters).

This set of observations establishes that our device is indeed a 
cavity soliton laser. Note that, from an applicative point of
view, these experiments open up the all-optical control of
microlasers on demand.
%We mention that the experiment yields also first indications for
%interaction between these spots. It is difficult to see on the scale
%of Fig.~\ref{fig:nf_switchONOFF}, but readily apparent in the movie
%contained in the supporting material.
    % The spot already existing on
    %the lower left side of the spot written in
    %Fig.~\ref{fig:nf_switchONOFF}b changes slightly position after the
    %switch-on and returns to its original position after the switch-off
    %in Fig.~\ref{fig:nf_switchONOFF}h.

Recorded spectra showed that the frequency of a specific CS could 
vary between successive on-switches, and could also change 
significantly after removal of the WB. These
frequency variations can be interpreted on the basis that
there is a family of CS solutions, each associated with different
longitudinal modes of the external cavity. The finite frequency
selectivity of the feedback allows this family to span about a 2~GHz
range for a fixed current and temperature. This enables different
family members to be excited by the WB, and permits noise-driven
jumps between them. Thus there is no
`memory' of the WB frequency, which is therefore only activating
CS lasing, not determining the CS properties.

\begin{figure}[ht!]
\centerline{
     \includegraphics[scale=0.31,angle=0]{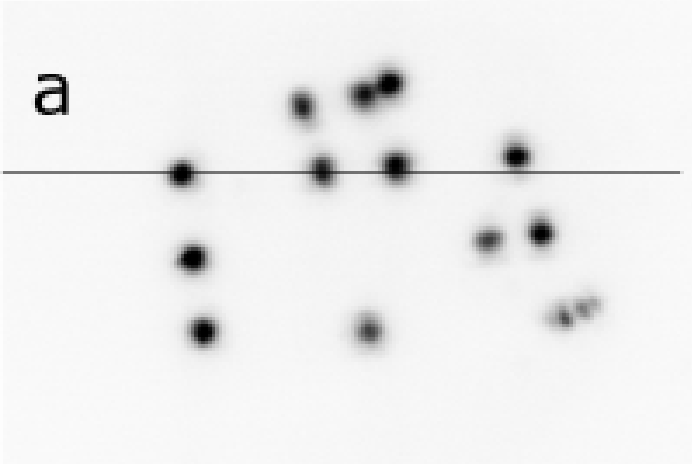}
     \includegraphics[scale=0.31,angle=0]{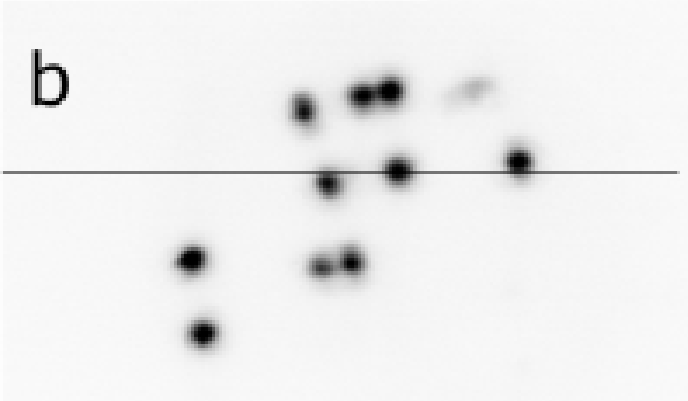}
     \includegraphics[scale=0.31,angle=0]{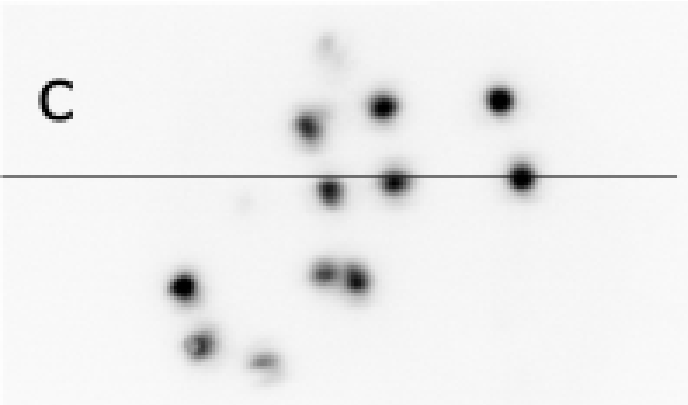}
     \includegraphics[scale=0.31,angle=0]{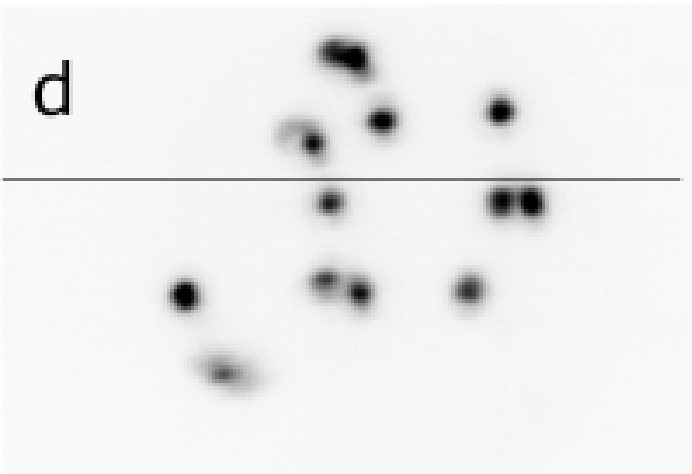}
     }
\caption{Near-field displaying the effect of a comb filter placed in
the external cavity at the re-imaged near-field position. Dark areas
correspond to high intensities. This filter allows feedback only on
several horizontal stripes with width $16~\mu$m at the VCSEL, spaced
by the same distance. From a) to d) the comb filter is moved
downwards. A line is added for reference, clearly showing that CS
positions are shifted between panels. For example the rightmost CS
close to the line in a) shifts quasi-continuously by about $18~\mu$m
between a) and d). Some CSs appear and disappear spontaneously, as a
result of local inhomogeneities. } \label{fig:nf_comb}
\end{figure}

Fig.~\ref{fig:LI} shows that some emission is already present
 (above 452~mA) before the CS actually switches on.
This emission comes from low amplitude background states or
patterns, visible also in Fig.~\ref{fig:nf_switchONOFF}. These are
blue-detuned by some tens of GHz to the CSs and their linewidth is
considerably larger. Their excitation is attributed to the finite
frequency selectivity of the system. This optical background seems
to be detrimental to the existence of the CSs. In order to reduce
it, a comb filter was inserted in the external Littrow cavity,
positioned as close as possible to the re-imaged near-field at the
grating (Fig. \ref{fig:setup}). This filter allows feedback only on
several horizontal stripes, and thus creates a near one-dimensional
CS confinement. Figure~\ref{fig:nf_comb} shows the resulting
near-field with this filter, where the optical background is clearly
weakened and more CSs have emerged. The characteristic in
Fig.~\ref{fig:LI}, taken with a comb filter, shows an extended
region at lower current where there is coexistence of the CS and a
background-free state (the non-lasing off-state of the VCSEL).

    %We mention that using a Littman setup is also beneficial in reducing
    %the optical background (even without the comb filter) due to the
    %increased frequency selectivity.

Further evidence for motility of CSs is shown by moving the comb
filter vertically (see the other panels in Fig.~\ref{fig:nf_comb}).
The CSs react by a change of position. Within some areas of the
sample, their shift is quasi-continuous, in others they seem to
`jump' from one preferred location to the other.
    %We also demonstrated motility by using a perturbation by the WB,
    %i.e., attracting spots with an injected beam and displacing them
    %over several diameters by moving the beam around (not shown).
    %The above results demonstrate that {\it the lasing spots have all
    %the properties of CS} \cite{barland02}, and can be named as such.
Their motility distinguishes CS from fundamental modes of bistable
small-area laser defined, e.g., by micro-machining. Its utilization
enables new features as all-optical delay lines, optically
controlled beam steering and self-alignment to different kinds of
external inputs.

The experiments shown in Fig.~\ref{fig:nf_comb} reveal also (as do
the images in Fig.~\ref{fig:nf_switchONOFF}) that the CS do not have
complete freedom of location in the present device, but tend to
be attracted to, and trapped by, certain `defects'.  Trapping
effects were also found in other CS systems based on similar VCSELs
\cite{barland02,hachair06}. These defects are no doubt
due to local inhomogeneities in the active layer (temperature or
carrier density inhomogeneities) and/or the mirror layers (index
inhomogeneities). The sensitivity of the CS location to such
imperfections suggests a diagnostic application, i.e. that
operation of a VCSEL in CSL mode can reveal structural defects
not amenable to conventional microscopy techniques.

There is a novel and interesting self-induced force acting on 
the CS in our system. The GHz spread of CS frequencies means the
emission frequency cannot be locked to the grating's peak feedback 
frequency (see also \cite{naumenko06}). Any detuned beam is fed 
back at a slightly different angle from its emission, which induces 
a phase gradient (of 2.9~mrad/$\mu$m/GHz) across the CS. Phase 
gradients are well known from other CS systems to exert forces, leading 
to CS drift in an otherwise homogeneous system \cite{firth96}. This means
that the preferred CS locations in our system are actually equilibria
between attracting defect forces and grating-induced tilt forces. 
We have confirmed the presence of a significant tilt-force by reversing 
the orientation of the grating, whioh leads to small shifts in all the 
preferred CS locations. The directions of these shifts are consistent 
with the tilt-force model, and their signs indicate that all the CS are 
blue-detuned from the grating peak, i.e. between the 
grating and VCSEL frequencies.

We observe a scatter, in the range of few mrad, in the 
centers of the far-field intensity distributions of different CS,
consistent with equilibrium between defect and tilt forces being attained 
with tilts of a few mrad. This tilt is much smaller than the angular 
acceptance of the VCSEL resonance (see above), and so the effect on 
feedback efficiency is small. Direct experimental measurement of the tilt is quite difficult, however, because it seems to be much smaller 
than the angular far-field width of a CS, typically 44~mrad. 
%%
%%We do not have an independent measure of the exact location of the 
%% optical axis in the far field, but estimate it be within 10~mrad 
%%of the centroid of the intensity distribution. 
%%
%%%%%% not sure this is useful information...
%%

The switching behavior shown in Fig.~\ref{fig:nf_switchONOFF} is
also consistent with the defect model including the grating-induced
force. Switch-on works well with the WB directly on the target
location. For switch-off the WB is positioned to the side, so
as to perturb the trap and initiate erasure, probably via carrier
effects. However switch-off is not possible for every orientation
of the WB with respect to the CS. When the phase gradient is oriented
so as to direct the CS towards the trap, the CS reappears after
the WB is removed. At these orientations switch-on is possible,
however. Reversing the grating, and thus reversing the direction
of the phase gradient, swaps these locations to the other
side of the CS.

The switching sequence shown in Fig.~\ref{fig:nf_switchONOFF} was
performed with quasi-cw beams (i.e.\ by mechanically opening and
closing the beam path of the WB for some seconds) because in that
case the injection is easily visualized. However, switch-on and
switch-off are also possible with a pulsed WB. The minimum pulse
length investigated is 25~ns, with switch delays in the 10~ns range,
and is limited by the available acousto-optic modulator
\cite{tanguy07a}. Polarization and frequency-detuning between CS and
WB are not critical but will influence the necessary switching
energy and the resulting time delay. Switching off with a pulse
aimed directly on the CS location is also possible, and works
reliably.

In conclusion, we have demonstrated cavity-soliton microlasers in a
semiconductor-based system and their all-optical control.
    %The solitonic character of the lasing spots is evidenced by their
    %exhibiting properties such as bistability and independent
    %manipulation (writing and erasure) of several CS.
    %We also demonstrated their motility by displacing with an injected beam or
    %apertures.
The  control is incoherent, i.e., there are no stringent phase or
frequency requirements on the external beams used for writing,
erasing or manipulation of the CS. This makes the scheme robust for
applications. We anticipate that switching times are limited both by
the external cavity round-trip time and by carrier relaxation. The
latter could be improved by engineering of the carrier
lifetime~\cite{garnache02,avrutin00}. With regard to round-trip
time, we have preliminary evidence of CS in a short cavity closed by
a volume Bragg grating \cite{radwell07u}. Thus there is every
prospect that use of microoptics (potentially to the level of
monolithic integration) will allow significant improvements in both
speed and compactness of these devices.

\section*{Acknowledgements}
This work was supported by the European Union within the FunFACS
project and by the Faculty of Science of the University of
Strathclyde with a starter grant. We are grateful to FunFACS
partners and to P.~Paulau, N.~A.~Loiko, and A.~V.~Naumenko for many
useful discussions and suggestions. M.~Sondermann and  F.~Marino
contributed to preliminary experimental stages of this work,
discussions with J.~R.\ Tredicce stipulated it.

\bibliography{biblio}

\end{document}